\documentclass{evn2004}
\usepackage{txfonts}
\usepackage{graphicx,graphics}
\usepackage{epsf}

\begin{document}

\title{VSI-E Software Suite}
\author{D. Lapsley\inst{1} and A. Whitney\inst{1}}
\institute{MIT Haystack Observatory, Off Route 40, Westford, MA 01886, USA}

\abstract{
As broadband access to high speed research and education networks has 
become increasingly available to radio telescopes around the world
the use of e-VLBI has also increased. High bandwidth e-VLBI experiments
have been achieved across wide areas [\cite{bib-gigevlbi}]. e-VLBI has
also been used for the transfer of data from ``production'' experiments.
As the use of e-VLBI becomes more and more prevalent, the need to
have some form of standard framework for facilitating interoperability
between various acquisition, transport and processing systems around
the world is becoming increasingly important.
This is the motivation behind the VLBI Standard Interface -- Electronic (VSI-E)
standard. VSI-E is currently in draft form [\cite{bib-vsie}] and is going
through the standards process within the VLBI community. In this
poster, we describe an initial reference implementation of the VSI-E protocol.
The implementation has been done using the C/C++ language and is in the
form of a re-useable library. It has been developed on the Linux platform
and can be easily ported to most POSIX-compliant platforms. The reference implementation also includes a high-performance application level transport protocol.  That can be used to transport data within the framework of VSI-E.
}

\maketitle

%

\section{Introduction}

As the use of e-VLBI becomes more and more prevalent, so the 
importance of a standard framework for the transport of VLBI
data across wide area networks and between various data acquisition
and data processing systems increases.

VSI-E was first discussed in detail at the Second e-VLBI Workshop in
Dwingeloo, The Netherlands [\cite{bib-dwingeloo}] in May 2003.
A draft VSI-E specification was subsequently distributed in early
2004 [\cite{bib-vsie}]. The main objectives of VSI-E are:

\begin{enumerate}
\item \emph{Interoperability:} through a common data format.
\item \emph{``Internet Friendliness'':} through the use of protocols that are
well known and well understood throughout the Internet community.
\item \emph{Ease of Implementation:} through the use of existing and/or
new libraries.
\item \emph{Transport flexibility:} through the use of a framework that will
allow users to choose their transport mechanism/protocol to suit their
network and/or throughput requirements.
\end{enumerate}

For this reason, the draft specification[\cite{bib-vsie}] proposes the use of extensions to the Internet Engineering Task Force (IETF) Real-time Transport Protocol (RTP)[\cite{bib-rtp}] suite of protocols for the transport of e-VLBI data.
RTP is able to meet all of the objectives above.

In this poster, we provide a brief description of the VSI-E protocol. We then
describe a reference implementation of the VSI-E protocol that is available
to the VLBI community[\cite{bib-vsieimpl}].

\section{VSI-E}

VSI-E makes use of the RTP suite of protocols[\cite{bib-rtp}]. This includes two protocols:

\begin{enumerate}
\item \emph{RTP:} which provides a means for encapsulating real-time data streams and transporting them across Wide Area Networks while maintaining timing synchronization. 
\item \emph{The Real-time Transport Control Protocol (RTCP):} which provides a control channel for RTP streams that is used to exchange management information as well as sender/receiver-side statistics and timing synchronization information.
\end{enumerate}

RTP has been used within the Internet community for many years to transport real-time data streams. Because of this, there is an active community of users, and there is a wealth of information about the implementation and operation of RTP that can be drawn upon. 
RTP has been designed with scalability and extensibility in mind. It is considered ``Internet friendly'' within the Internet community and can be easily extended to accomodate e-VLBI requirements.
For more information on the use of RTP to transport VLBI data, refer to the VSI-E draft specification [\cite{bib-vsie}].

\section{VLBI Real-time Transport Library}

In this section we discuss a reference implementation of the draft VSI-E specification described in the previous sections. Due to space limitations, we will only describe highlights of the system. This implementation was initially done using C/C++ on a Linux Redhat 9.0 system. It is easily portable to most POSIX-compliant UNIX systems.

The main design goals of this implementation were:
\begin{enumerate}
\item \emph{Performance:} targeted transfer rates as close to line rate as possible.
\item \emph{Ease-of-use:} C++ classes that encapsulated large amounts of functionality and provided a simple, high-level interface that would give users the capability to create powerful programs with minimum effort.
\item \emph{Flexibility:} provide the capability to use different transport mechanisms/protocols for input/output
\end{enumerate}

The library developed provides a number of C++ classes (or software modules) that encapsulate various system and protocol-level functionality:
\begin{enumerate}
\item \emph{Mutex} and \emph{Guard} classes encapsulate Mutual Exclusion system calls used to serialize concurrent access of data from multiple threads.
\item A \emph{Thread} class encapsulates POSIX threads.
\item \emph{Socket} and derived classes encapsulate TCP, UDP, Unix Domain and VLBI Real-time socket functionality using the same programming interface. A FileSocket is also included in this family of classes to provide a uniform access interface.
\item \emph{RTCPSocket} and derived classes encapsulate RTCP control message transfers.
\item \emph{RTPSocket} and derived classes encapsulate RTP data message transfers.
\item An \emph{RTPSession} class encapsulates an RTP session's state information.
\item \emph{VRSocket} and derived classes encapsulate application layer functionality and interactions. 
\end{enumerate}

\begin{figure}[t]
	\centering
   \includegraphics[width=8cm]{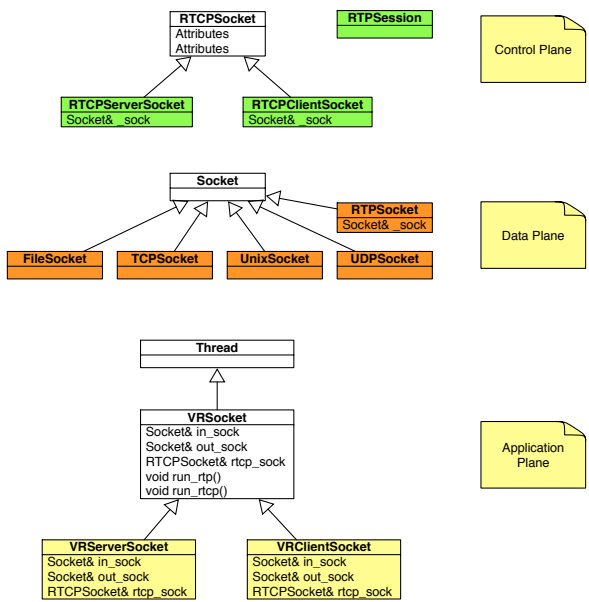}
   \caption{Class Hierarchy}
   \label{fig:hierarchy}
\end{figure}

Figure~\ref{fig:hierarchy} shows a cut-down UML diagram of a portion of the library's C++ class\footnote{For reader's unfamiliar with C++: a ``class'' can be considered a software module that encapsulates data and the operations performed on data in a manner that is easy to use and integrate with other ``classes''; ``inheritance'' is a technique that allows a ``child'' class to ``inherit'' all of the functionality of a ``parent'' class}.  hiearchy. Base classes are in white and implement functionality common to a group of classes. Classes that can be used by users are shaded. The classes can be divided into three levels: control plane, data plane and application plane.
In order to use the library, a user would normally start by creating connections between two machines through the use of data plane objects (e.g. TCPSocket). Typically, two connections would be created (one for RTP data and the other for RTCP control messages). Then RTCPSocket objects and RTPSocket objects would be created on top of the data plane messages (note that an RTPSocket is at a higher level in the data plane). Finally, the application layer objects VRServerSocket and VRClientSocket could be built on top of the RTPSocket and RTCPSocket. At this point, the user would call the rtp\_proc() and rtcp\_proc() methods of the VRServerSocket and VRClientSocket objects to start data and control message transfer.
This model provides significant programming flexibility for the user. In particular, it allows the user to choose the order in which connections are setup and what transport protocols are used to transport data. For example, a user could create a program that used TCP to transport RTCP control messages and TCP to transport RTP data across a shared IP network. Another user may have a dedicated network and wish to use UDP instead. This can be done by changing a few lines of code. Yet another user may wish to use a proprietary, high-performance protocol of their own design to transport data across shared networks. The library has been designed to allow users to extend it in such a manner seamlessly.

The library has also been used to build an application that is available for people to use to transfer data without any programming effort.


\section{Summary}

In this poster, we have discussed the increasing importance of VSI-E as the use and extent of e-VLBI continues to increase.
We have briefly described the objectives of VSI-E and the current draft VSI-E specification.
We have also described a reference implementation of the draft specification that provides a flexible, powerful framework that can be used to construct VSI-E compatible applications.


\end{document}